\documentclass[preprint aps, amsmath, amssymb]{revtex4-2}

\usepackage{graphicx} 
\usepackage{color}
\usepackage[normalem]{ulem}
\usepackage{booktabs}
\usepackage{makecell}
\usepackage{hyperref}

\begin{document}
\title{Writing and readout of luminescent defects on an optical nanofiber using an electron beam}
\author{Yining Xuan}
\author{Mark Sadgrove}
\email{mark.sadgrove@rs.tus.ac.jp}

\address{Department of Physics, Tokyo University of Science, Tokyo 162-8601, Japan\\}

\begin{abstract}
We demonstrate the writing and readout of groups of luminescent defects on an optical nanofiber (ONF), with readout performed through the fiber modes.
In the write step, a focused electron beam incident on the fiber surface created luminescent defect centers in the fiber's silica in a region of diameter $\sim200$ nm localized about the beam center. In the readout step, the electron beam was scanned over the fiber and the induced cathodoluminescence from the defects was coupled into the ONF and detected. The regions with electron beam-induced luminescent defects exhibited a strongly enhanced optical signal, despite minimal observable change in the fiber surface, as judged by secondary electron imaging over the same region. Spectral measurements show that the enhanced luminescence stems from the creation of both oxygen deficiency centers (ODCs) and non-bridging oxygen hole centers (NBOHCs) by the electron beam. 
\end{abstract}
\maketitle
\section{Introduction}

Optical nanofibers (ONFs) are optical fibers with diameters comparable to or smaller than the optical wavelength of propagating light. They are known to provide strong transverse confinement of light and efficient coupling between quantum emitters and guided modes via their relatively strong evanescent field~\cite{Tong2003,Balykin2004}. Because of these properties, ONFs have become versatile platforms for nanophotonics, quantum optics, and remote optical sensing \cite{Nayak2018}. 

Although studies of ONF-coupled photon sources up to now have focused on coupling from external emitters~\cite{aharonovich2016solid} or emitters doped into the fiber core~\cite{Shimizu2024}, silica itself is also known to host a variety of luminescent defect centers. Among them, the oxygen defect center (ODC), which emits near 460 nm, and the non-bridging oxygen hole center (NBOHC) which emits near 650 nm are the most extensively studied \cite{griscom1985defect,friebele1985optical,Tohmon1989,Skuja1998,Suzuki2003,girard2019overview}. 
Furthermore, cathodoluminescence (CL) studies of silica-based materials have shown that electron irradiation can generate or activate defect-related emission bands in the visible range \cite{Fitting2005,girard2019overview}. More recently, NBOHC-related luminescence has also been exploited for measurement of the photonic density of states of a thin fiber~\cite{Uemura2021} and for high-resolution optical imaging in electron-irradiated silica-containing systems \cite{Hung2025}.
The concept of using electron beam radiation to create localized luminescent centers has also been tested in other materials such as halide perovskite ~\cite{fujimaru2026precisely} and hexagonal boron nitride~\cite{roux2022cathodoluminescence}.

The main motivations for studying the controlled creation of defect centers in silica are twofold. 
First, such defects are closely related to the optical properties of silica, yet their behavior in strongly confining nanophotonic geometries such as ONFs remains only partly explored. 
Second, if the number and position of the generated defects can be controlled at the single-defect level, they may provide a new route toward fiber-integrated quantum emitters. 

In this work, we use electron beam irradiation to create groups of luminescent defect centers near the surface of an ONF, which can then be read out as CL through the fiber mode itself by performing an electron beam scan over the area which was written to. 
Although the present work deals with ensembles of defects rather than isolated emitters, it provides a first step towards understanding whether single electron-beam-induced luminescent centers can be created, addressed, and read out in an ONF prepared from a standard commercial optical fiber.

\section{Experimental principles}

\subsection{Writing: Defects in silica and their creation by the electron beam}
Due to their effect on the propagation of light in optical fibers, luminescent defect centers in silica have been 
well studied over the past decades~\cite{griscom1985defect,friebele1985optical,Tohmon1989,Skuja1998,Suzuki2003,girard2019overview}, with Ref.~\cite{Skuja1998} providing a convenient reference. Among the known defect centers, we focus here on the two mentioned in the introduction: the ODC of type II
(emission peak $\sim460$ nm) and the NBOHC (emission peak $\sim650$ nm) as the observed spectra in our experiment correspond to the known photoluminescence bands of these defects. Note that NBOHCs are considered to be oxygen \emph{excess} centers, in contrast with ODCs. There have been a number of studies on the creation of these defects due to radiation~\cite{friebele1985optical,girard2019overview} including by electron beam radiation~\cite{sergeev2004electron,Fitting2005,vaccaro2007radiation,girard2019overview,Hung2025}.
On the other hand, defects occur in the silica matrix of optical fibers 
without any particular need for irradiation~\cite{friebele1985optical}. 
ODCs are considered to be the more plentiful defect in bulk silica~\cite{vaccaro2007radiation}.

We also note that ODCs and NBOHCs have similar oscillator strengths, but very different excited state lifetimes. In particular, Skuja gives
a lifetime of $\sim 10$ ms for the ODCs observed in our experiment, and a comparatively short lifetime of $\sim 15$ $\mu$s for NBOHCs~\cite{Skuja1998}.
However, both of these lifetimes are long compared to those of solid state quantum emitters typically used in single photon generation experiments~\cite{yu2025concise}.

\begin{figure}[htbp]
\centering
\includegraphics[width=12cm]{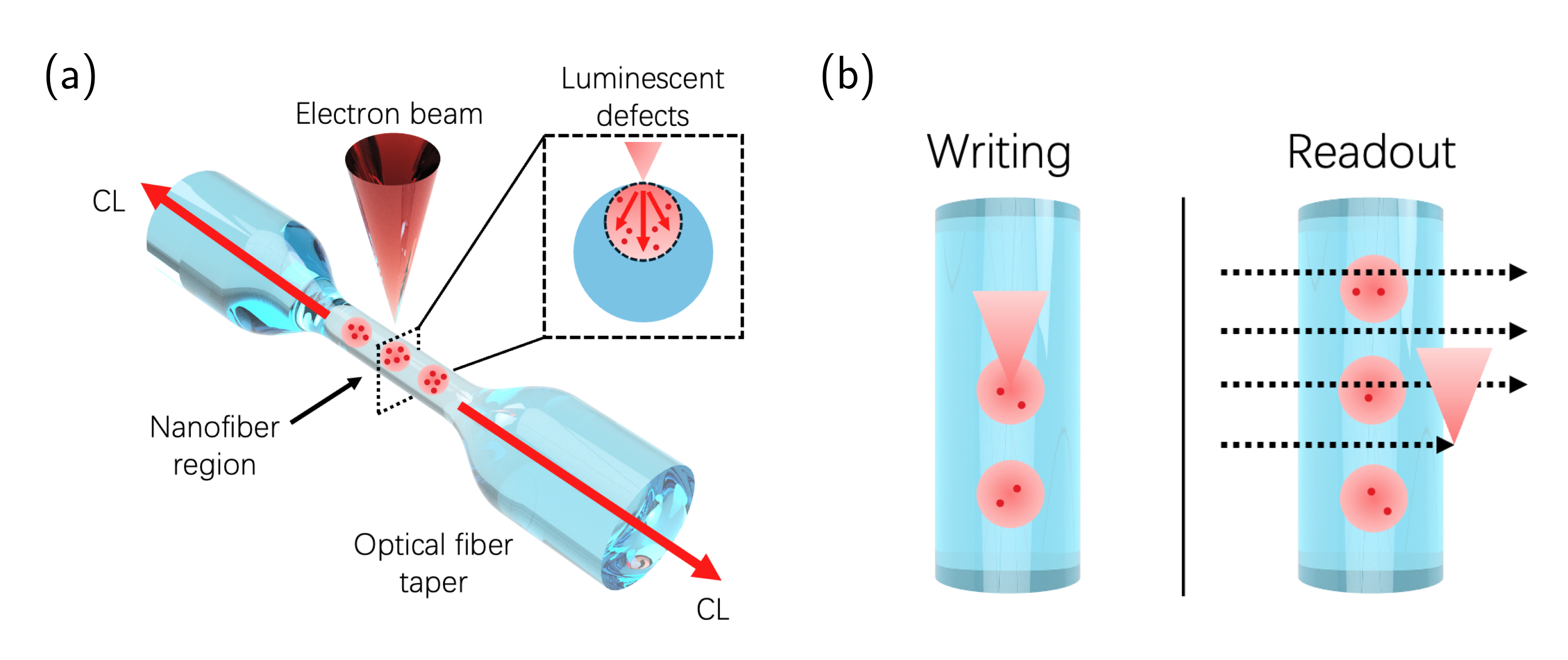}
\caption{\label{fig:exp}
(a) Concept of the experiment. An electron beam is used to write or readout luminescent defects on an optical nanofiber. The inset shows a cross-section of the fiber, and depicts the creation of a volume including luminescent defects below the fiber surface.
(b) writing and readout process. The focused electron beam is operated in spot mode for writing, and returned to scan mode for readout.}
\end{figure}

Figure~\ref{fig:exp}(a) shows the concept of our experiment. The inset shows how the electron beam produces an interaction volume near the fiber surface~\cite{goldstein2018scanning} in which the incident electron along with back-scattered electrons can provide the energy to produce luminescent defect centers. We estimate that the diameter of the interaction volume is about 200 nm for the parameters used in our experiment, or about 30 times the diameter of the electron beam.

\subsection{Readout: Nanofiber-coupled cathodoluminescence}

In order to observe defect centers in the ONF's silica matrix, we collected the incoherent CL~\cite{garcia2010optical,polman2019electron} from the defect centers induced by the electron beam.
The collection of CL was performed through the modes of the nanofiber itself, which the defect centers near the fiber surface couple to with an efficiency of approximately 24$\%$, assuming random polarization, as estimated by finite-difference time-domain simulations~\cite{yalla2012fluorescence, yalla2012efficient}.

We note that such waveguide-collected CL for emitters on or inside the fiber is a fairly recently developed technique~\cite{Uemura2021, muller2021broadband,scheucher2022discrimination,arend2025electrons,xuan2026creation}, although collection of CL via a fiber facet placed near to a sample has been used for some time~\cite{hocker2017three}. Here, we utilize ONFs as both the medium on which luminescent defects are written, and also as the waveguide through which their electron-beam induced luminescence (i.e. cathodoluminescence) is read out.

Let us briefly mention some unique aspects of our nanofiber-based CL readout technique. First, the CL image is reconstructed from asynchronous intensity measurements using the fact that the CL signal only exists when the electron beam passes over the fiber to  combine these one dimensional traces into a single CL image. In particular, the sharp signal increase at the fiber edge allows the 
signals to be appropriately aligned. Second, one non-ideal aspect of the method is that the fiber, being suspended, can move slightly during the scan, which takes on the order of several minutes to complete. This can be compensated for in our offline data analysis.
More details regarding our method may be found in Refs.~\cite{Uemura2021,xuan2026creation}, and the code used to reconstruct CL images is available as a FigShare archive~\cite{figshare}.

\section{Experimental setup}

We now move on to details of the experiment. ONFs were fabricated from a commercial single-mode silica fiber (780HP) by a heat-and-pull tapering process~\cite{birks1992shape,graf2009fiber,ward2014contributed}. Use of a constant width hot zone during heat-and-pull produced a taper with an exponential profile and a fiber waist with a minimum diameter of approximately 500~nm, and a length of approximately 1~mm. Because suspended ONFs are mechanically fragile and sensitive to beam-induced vibration during SEM measurements, we selected a relatively thicker section of the taper, with a diameter of up to 900~nm, for the present experiments, although reducing the electron beam current should in principle allow writing on thinner sections of the fiber.

The ONF was mounted inside the chamber of a field-emission SEM (Carl Zeiss SUPRA 50). The ends of the fiber were fusion spliced to a homemade feedthrough~\cite{abraham1998teflon} allowing photons coupled into the guided modes of the ONF to be measured outside the chamber.

The experiment consisted of two steps, namely writing and readout. In the write step, depicted in Fig.1(b), the SEM beam was operated in spot mode rather than in normal raster-scan imaging mode. 
The focused electron beam was positioned on the ONF surface at the position where creation of a region of luminescent defects was desired.
Note that because the beam was stationary during this step, conventional SEM observation of the region where writing occurred was not possible simultaneously with writing. 

In the readout step, as depicted in Fig.1(b), the beam was returned to scan mode and rastered over the ONF region. The resulting CL, coupled into the guided modes of the ONF, propagated to both fiber ends and was detected outside the chamber using single-photon counting modules (SPCMs, Excelitas SPCM-AQRH-15-FC). The photon counts were recorded and processed offline to reconstruct the CL image.

To investigate the origin of the observed CL enhancement, we also performed spectral measurements using an optical multichannel analyzer (OMA, Kymera 193i, Newton DU970P-BVF). For this experiment, a broader region of the ONF was exposed by the electron beam. Spectra from non-exposed and exposed regions were then measured over the wavelength range from 250 to 1000~nm.

\section{Results}

\subsection{Basic properties of writing and read-out}

\subsubsection{Spatial properties of luminescent defect regions}

We first wrote luminescent defect regions onto the fiber surface at a position where the diameter of the fiber was approximately 900 nm. At an acceleration voltage of 2~kV, the electron beam diameter (2$\sigma$ diameter) was approximately 6.4~nm, and the electron beam current was $\sim 200$ pA~\cite{Uemura2021}.

\begin{figure}[htbp]
\centering
\includegraphics[width=12cm]{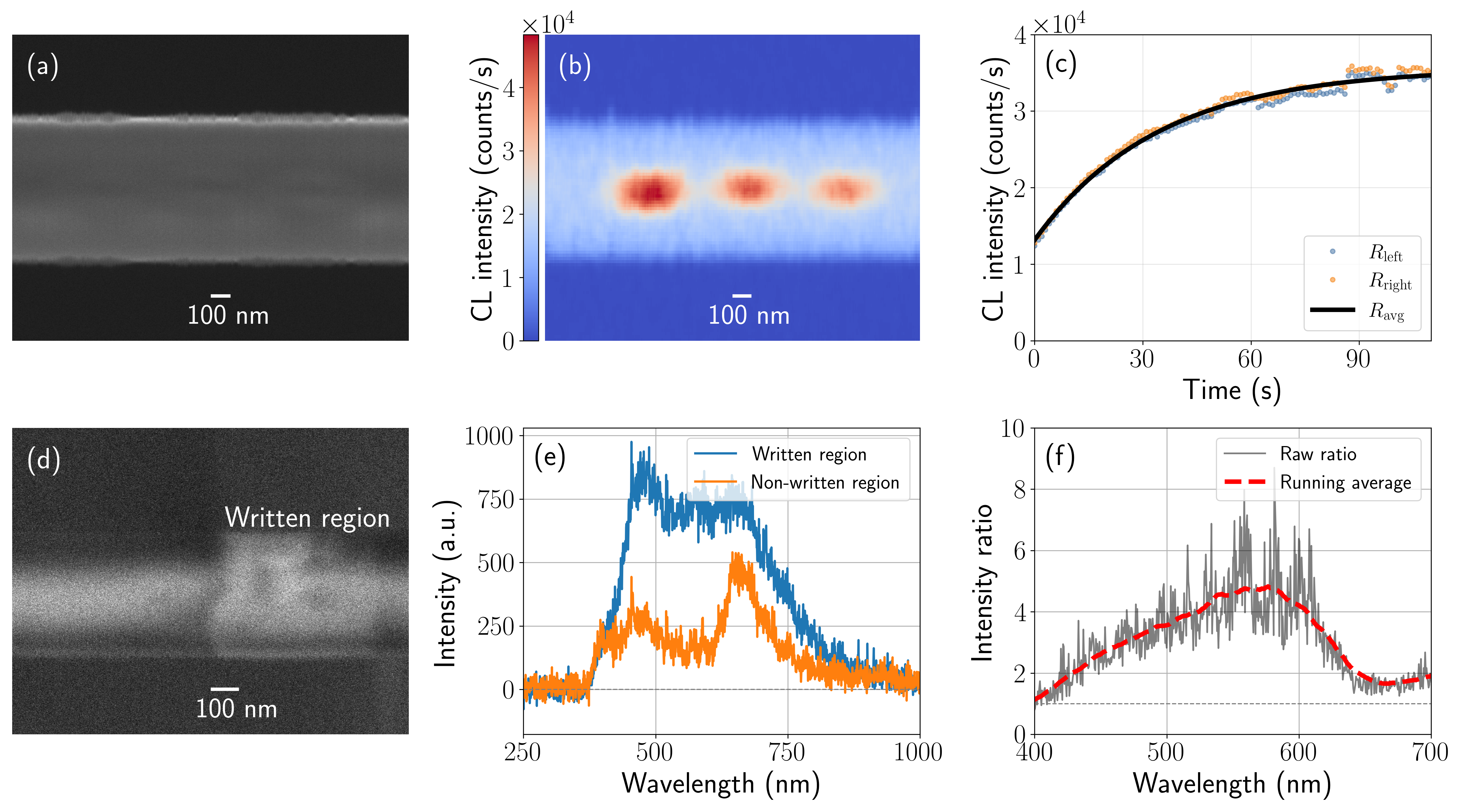}
\caption{\label{fig:write}Writing and readout of luminescent defect regions along with spectral characterization.
(a) SEM image after three defect regions have been written.
(b) CL signal from fiber after writing of the three defect regions. 
The duration of the write operation was (from left to right): 90s, 60s, 30s.
(c) Temporal variation of the CL signal under constant illumination. The blue and orange data points represent $R_{\rm left}$ and $R_{\rm right}$, respectively, while the black solid line shows the exponential fit to the average of the left and right signals $R_{\rm avg}$.
(d) SEM image showing the written region used for spectral measurements.
(e) CL spectra measured from the written and non-written regions of the ONF, as indicated in the legend.
(f) Intensity ratio between the written and non-written spectra in the visible range. The red dashed line shows the running average of the ratio.}
\end{figure}

To compare the results obtained for different illumination times, we irradiated equally spaced positions on the ONF surface in the SEM spot mode for exposure time $t_{\rm exp}$ of 30~s, 60~s, and 90~s, respectively. The standard SEM image and CL image of the sample after writing are shown in Figs.~2(a) and 2(b) respectively.
In the SEM image in Fig.2(a), the three irradiated positions are difficult to identify clearly.
By contrast, in the CL signal shown in Fig.2(b), the three luminescent defect regions can be clearly observed. 

The radius $\sigma_x$ of the regions in the horizontal direction and that $\sigma_y$ in the vertical direction was seen to only slightly increase with increasing illumination time, having a value of about 100 nm in all cases. Note that this is much larger than the radius of the electron beam itself, due to the interaction volume of incident electrons with the silica of the fiber~\cite{goldstein2018scanning}. On the other hand, the maximum count rate $R_{\rm peak}$, was found to increase with illumination time as summarized in Table 1.

Fig.2(c) shows the measured temporal variation of the CL signal under constant electron beam illumination. 
The electron beam was focused onto the ONF surface in spot mode, and the CL signal intensity was 
recorded as a function of time. The count rates measured from the left and right detection channels were first averaged to give

\begin{equation}
R_{\rm avg}(t)=\frac{R_{\rm left}(t)+R_{\rm right}(t)}{2}.
\end{equation}

The averaged time trace was then fitted using the exponential function

\begin{equation}
R_{\rm avg}(t)=R_{\rm sat}-(R_{\rm sat}-R_0)
\exp\left(-\frac{t}{T}\right),
\end{equation}

where $R_0$, $R_{\rm sat}$, and $T$ are the initial count rate,
the saturated count rate, and the characteristic saturation time, respectively.

The fitted initial count rate is $R_0 = 1.316\times10^4~{\rm counts/s}$, and the saturated count rate is $R_{\rm sat} = 3.561\times10^4~{\rm counts/s}$.
The fitted time constant is $T=34.50~{\rm s}$. 

After $\sim 100$ s of illumination, we typically noticed variations in the signal which we assume were due to fiber vibrations induced by the mechanical 
effect of the electron beam scattering along with local heating of the fiber.

\begin{table}[t]
\centering
\caption{Measured parameters of defect regions.}
\label{tab:gaussian_fit_results}
\begin{tabular}{lcccc}
\hline
Region &
$t_{\rm exp}$ &
$R_{\rm peak}$ &
$\sigma_x$ &
$\sigma_y$ \\
&
(s) &
(counts/s) &
(nm) &
(nm) \\
\hline
Left   & 90 & $4.84\times10^4$  & $120$ & 100  \\
Middle & 60 & $4.38\times10^4$  & $89$  & 86   \\
Right  & 30 & $3.94\times10^4$  & $99$  & 83  \\
\hline
\end{tabular}%
\end{table}

\subsubsection{Spectral properties of the luminescent defects}

To examine the origin of the enhanced optical signal, we compared the CL spectra from non-written and written regions of the ONF. (See Fig.~\ref{fig:write}(d)). We excited only a region with an area of a few hundred square nanometers using the SEM electron beam, while one end of the ONF was connected to the OMA for spectral measurements.

Fig.~\ref{fig:write}(e) shows that the written region exhibits stronger emission over a broad wavelength range associated with ODC and NBOHC defect centers. Fig.~\ref{fig:write}(f) shows the intensity ratio between the written and non-written regions in the 400--700~nm range. The red dashed curve is obtained by applying a 100 points moving average window to the raw spectral intensity ratio. It may be seen that the wavelength region associated with the ODC emission band is slightly more intense than that associated with the NBOHC emission band, suggesting that more ODCs than NBOHCs are produced by the writing process.

\subsection{Luminescent pattern writing using beam spot translation}

\begin{figure}[htbp]
\centering
\includegraphics[width=12cm]{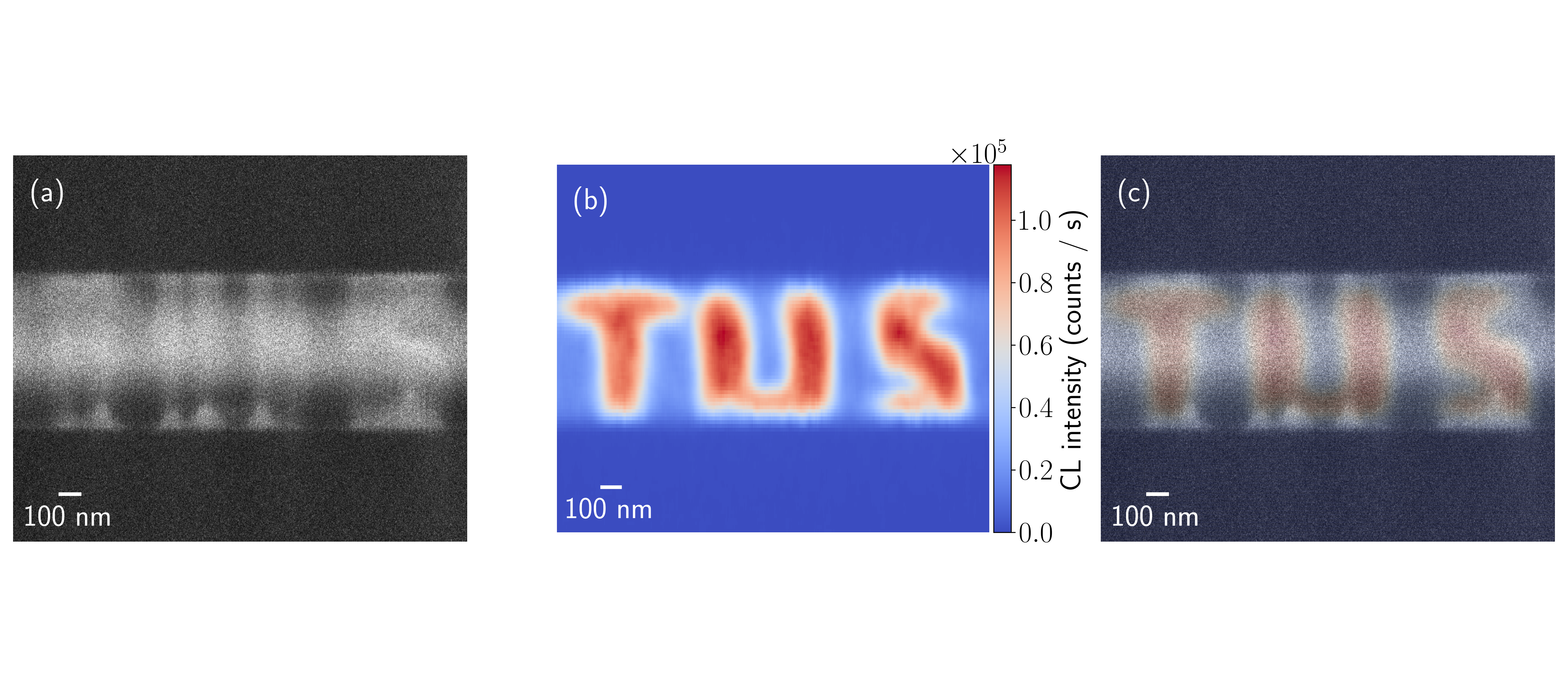}
\caption{\label{fig:3}Writing and readout of specific patterns on the ONF.
(a) SEM image of the nanofiber after writing. (b) CL image of the nanofiber after writing.
(c) Overlay of (a) SEM image and (b) CL image.}
\end{figure}

As shown above, defect regions can be written freely at a given position using the SEM spot mode. 
However, it is also possible to move the spot in real time, allowing patterns to be written. 
We show an example of this in Fig.~\ref{fig:3}.

After writing, the post-exposure SEM image Fig.~\ref{fig:3}(a) is visibly altered, but it is not obvious what the luminescent defect pattern is from this secondary electron image alone. Although it is difficult to confirm the precise cause of the changes seen to the fiber surface due to exposure to the electron beam spot, we expect that it arises from a combination of charge up and modification of the silica surface due to heating, and the creation of the defect centers themselves. 

The waveguide-collected CL image shown in Fig.3(b) reveals clearly the pattern which has been written (in this case the Tokyo University of Science logo "TUS"). 
The overlay image in Fig.3(c) further shows a close spatial correspondence between the beam-affected regions visible in the SEM image and the high-intensity regions in the CL image. 

\section{Discussion}

\subsection{Lower bound on the number of the luminescent defects}

In experiments where optical excitation of defect centers can be performed, the density of defects can be estimated from photoluminescence measurements and knowledge of the defects' absorption 
cross sections, along with other parameters, at the specific optical frequency used~\cite{messina2010generation}. 

Unfortunately, in our current experiment, optical excitation of the written defect centers was not possible, and, due to the lack of a detailed model for how electron irradiation excites the defect centers under consideration, a number of uncertainties exist in any estimate of the number of luminescent defects from our photon counting observations alone. 

One available strategy is to assume that the defect centers are effectively driven well above saturation intensity so that the number of photons each defect produces is limited only by the transition lifetime $\tau$. If the transition is not actually driven sufficiently above saturation to achieve this condition, an observed photon count rate of $1/\tau$ should instead be attributed to multiple emitters, and thus an estimate assuming excitation above saturation intensity should be considered to be a lower bound on the luminescent defect number. 

Before estimating this lower bound, let us comment on the plausibility of saturating the transition by electron beam illumination. First, we note that the broad spectra we see in both the ODC and NBOHC bands have the qualitative characteristics typically associated with off-resonant excitation along with a phonon-broadened spectral bandwidth. Thus, it seems reasonable to model the defect centers as three level systems, in which electron beam-induced off-resonant excitation allows rapid population of the excited state of the optical transition via phonon-mediated decay, and a life-time limited decay rate of the optical transition itself. Second, we note that the defect excited state lifetimes are much longer than the average time separation of electrons $\Delta t= e/I_{\rm beam}\sim1$ ns, where $I_{\rm beam}\approx 200$ pA is the electron beam current, meaning that the electron beam provides effectively continuous excitation. (We note that there exist a number of studies of photon bunching due to electron beam illumination, in which case the arrival of electrons should be considered as producing discrete excitation events relative to the scale of the transition lifetime~\cite{meuret2015photon,yanagimoto2023time,yanagimoto2025unveiling}, so that the continuous illumination approximation is not guaranteed in general).    
Third, we can estimate the maximum possible intensity of excitation $I_{\rm ex}$ by dividing the power $P = V_{\rm acc}I_{\rm beam}$ by the electron beam area, assuming a beam radius of 3.2 nm. Then, for an acceleration voltage $V_{\rm acc}=2$ kV, we find $I_{\rm ex}=1.2$ MW~cm$^{-2}$. We note that this is well above typical saturation intensities for quantum emitters in the solid state~\cite{novotny2012principles}. However, because not all of the electron energy is utilized in exciting the transition, it is difficult to relate this value directly to the excitation intensity. In addition, it is not clear what proportion of the luminescence is excited by the initial incident electron, and what proportion is due to consequent back-scattered electrons inside the silica.

Nonetheless, while the above calculations do not establish rigorously that the electron beam illumination saturates the optical transitions of ODCs and NBOHCs, they do not directly contradict the assumption either, and so we consider that it is meaningful to perform calculations in this limit.

We proceed to calculate the lower bound estimate of the luminescent defect number by assuming that the defects can be modeled as three level systems driven well above saturation, producing a photon count rate $R$ which can be written as
\begin{equation}
\label{eq:R}
    R = N \eta_{\rm fc}\eta_{\rm det}\frac{1}{\tau},
\end{equation}
where $N$ is the number of defects, $\eta_{\mathrm{fc}}$ is the overall probability that an emitted photon is coupled into the fiber fundamental mode, $\eta_{\mathrm{det}}$ is the detection efficiency at the emission wavelength, and $\tau$ is the optical transition lifetime. This formula also assumes that decay of the excited state of the optical transition always produces a photon, i.e., that the quantum efficiency is unity.

To estimate the relative contributions of the NBOHC and ODC(II) wavelength CL components, we measured the guided CL intensity using a 600~nm long-pass filter and a 600~nm short-pass filter, respectively. The long-pass and short-pass signals accounted for approximately 78.2\% and 17.9\% of the no-filter signal, respectively. The remaining difference, about 3.9\%, is likely attributable to the losses of the optical filters. Therefore, for this simple estimate, we assumed that the background-subtracted guided CL signal is shared by NBOHC and ODC(II) emission with the ratio of 4.37 : 1. In this case,

\begin{equation}
R_{\mathrm{NBOHC}} + R_{\mathrm{ODC}} = R.
\end{equation}

Assuming random emitter dipole orientation, we take the total coupling efficiency into the ONF guided modes to be $\eta_{\mathrm{ONF}}\sim0.24$ for 650~nm and 0.28 for 460~nm, respectively. The single-ended coupling efficiency is estimated as $\eta_{\mathrm{fc}}\sim\eta_{\mathrm{ONF}}/2$.
The detection efficiency is $\eta_{\mathrm{det}}=0.7$ for NBOHCs and $\eta_{\mathrm{det}}=0.25$ for ODCs.

Finally, using a representative room-temperature NBOHC lifetime of $\tau_{\rm NBOHC}\sim 14.5$~$\mu$s and an ODC(II) lifetime of order $\tau_{\mathrm{ODC}}\sim 10$~ms, the corresponding lower-bound estimates calculated from Eq.~(\ref{eq:R}) are
\begin{equation}
N_{\mathrm{NBOHC}} = 1.8,
\qquad
N_{\mathrm{ODC}} = 700,
\end{equation}
for $R=R_0$ and 
\begin{equation}
N_{\mathrm{NBOHC}} = 5,
\qquad
N_{\mathrm{ODC}} = 1900,
\end{equation}
for $R=R_{\rm sat}$.

Because studies exist where NBOHC concentrations have been measured for various silica samples, one way to check the validity of the above results is to convert them to a density and compare their values to those found in the literature. Here, we face two main problems. One is the wide range of NBOHC density values found in the literature, ranging from $10^{15}$ cm$^{-3}$ in fluorine doped optical fibers~\cite{vaccaro2012influence}, $\sim 10^{18}$ cm$^{-3}$ in UV bleached silica samples~\cite{skuja2011visible}, up to extremely high values similar to $10^{20}$ cm$^{-3}$ in radiation exposed samples~\cite{messina2010generation}. The other problem is that although we can estimate the volume in which NBOHCs are \emph{created} from our measurements shown in Fig.~\ref{fig:write}(b), it is not immediately clear if the volume in which NBOHCs are created is equivalent to the volume in which they are \emph{excited}, because writing is performed by a stationary electron beam, whereas the measurements in Fig.~\ref{fig:write}(b) necessarily come from raster-scanning of the electron beam over the entire defect region. That is, it is plausible that the volume in which defects are excited by the electron beam is \emph{smaller} than the volume in which they are created by the same beam. 

Because of this ambiguity, we will provide density estimates in both the case where the excitation volume is assumed to be the same as the volume in which defects are created and in the case where the excitation volume takes on its minimal plausible value, which is given by the cylindrical volume calculated from the electron beam radius and its penetration depth into the sample (175 nm in this case~\cite{raftari2018modified, Uemura2021}).

Taking $r=$100 nm in the first case, and assuming a spherical volume, we find the NBOHC density for the unirradiated sample to be $4.9\times10^{13}$ cm$^{-3}$. For the minimum plausible excitation volume,
the density becomes $4.9\times10^{17}$ cm$^{-3}$. While the former value is unreasonably low, the second value sits comfortably within ranges reported in the literature. 

We note that if number of luminescent defects is actually given by the lower bound, the number of NBOHCs in the excitation volume is predicted to be sufficiently small that weakly anti-bunched light should be measurable. The best strategy to achieve this would seem to be avoiding the creation of defects in excess of those already present in the fiber. To do this, the beam could be scanned rapidly over the sample at a low acceleration voltage to avoid defect writing. Even then, however, the beam size and beam current parameters would have to be adjusted to give the smallest interaction volume possible. This is a subject of on-going investigation, but we note that up to now we have not observed any signature of anti-bunched light. 

\subsection{Analysis of information density}

Although we do not intend for the effect demonstrated here to be used for data storage, etc,
it may be useful for some applications to consider the density of information stored in the defect regions.
From the measured center-to-center separations of the three defects in Fig.~2(b), which are approximately 0.43 and 0.52~\textmu m (average $\sim 0.48$~\textmu m), the demonstrated one-dimensional storage density is approximately

\begin{equation}
D_{\mathrm{1D}} \sim \frac{1}{0.48~\text{\textmu m}} \sim 2.1~\mathrm{bits/\text{\textmu m}},
\end{equation}

which corresponds to $\sim 2.1~$bits$/\mu$m. This already exceeds the linear density of a conventional compact disc ($1/0.833 \approx 1.2~\mathrm{bits/\text{\textmu m}}$), is close to that of DVD ($1/0.40 = 2.5~\mathrm{bits/\text{\textmu m}}$), but remains below Blu-ray ($1/0.149 \approx 6.7~\mathrm{bits/\text{\textmu m}}$) and far below modern LTO-9 tape ($\sim 21.5~\mathrm{bits/\text{\textmu m}}$) \cite{Sony2000,PioneerDVD,BluRay2018}. Because the present spacing was chosen for clear visual separation rather than for maximum packing density, this value should be regarded as a demonstrated density rather than an optimized limit. If the spacing were reduced toward the measured defect region diameter of $\sim 0.20~\text{\textmu m}$, the corresponding linear density would approach $\sim 5~\mathrm{bits/\text{\textmu m}}$, that is, of the same order as Blu-ray optical storage.

\section{Conclusion}
In conclusion, we have demonstrated electron-beam writing of luminescent defects near the surface of an ONF along with their readout through the same fiber. 
In the write step, the focused electron beam locally created luminescent defects in silica. 
In the readout step, raster scanning of the written region generated cathodoluminescence from the defects that coupled to the ONF guided modes. 
The shape of the written regions could not be clearly ascertained through a standard SEM measurement, but the reconstructed CL measurements exhibited strong optical contrast, allowing a clear image of the written region to be obtained. It is amusing to note that this property is somewhat analagous to the concept of "invisible ink" in that standard observation methods cannot make out the written characters, which must instead be revealed by a suitable "readout" process (e.g. by the application of ultraviolet light, etc, for standard invisible ink~\cite{macrakis2014prisoners}, and here by the application of the CL method). 

In addition spectral measurements showed that the luminescence from the written regions could be attributed to ODC- and NBOHC-related defects in silica. Lower bound estimates of the defect number suggest that the ODC defects are far more numerous (although they contribute only weakly to the optical signal), and that the number of written NBOHC defects created could be as few as $\sim 5$ in principle.

The results presented here should be viewed as an initial demonstration of the technique. The present implementation requires that the writing occur at a relatively thick portion of the taper waist region, although the diameter was still technically in the nanofiber regime. Mechanical stabilization of the fiber, along with optimization of the electron beam parameters to reduce heating and other fiber damage should allow writing to occur at thinner fiber diameters, improving the coupled optical signal intensity.
Additionally, estimates of the luminescent defect number suggest that the current method creates many defect centers simultaneously. Nonetheless, the results do suggest a route toward single emitter creation, and a natural next step would be to reduce the number of defects by optimizing the acceleration voltage and beam current, and correlating the written spot size and intensity with measured photon statistics. 

\section*{Acknowledgments}
This research was supported by a KAKENHI Grant-in-Aid for Transformative Research Areas (Grant No. JP22H05135), and the Nano-Quantum Information Research Division of Tokyo University of Science. The authors acknowledge informative conversations with Hikaru Saito.

\bibliography{main_reference}

\end{document}